\documentstyle[preprint,prd,aps,12pt]{revtex}

\preprint{
$
\begin{array}{r}
\text{LAVAL-PHY-96-16} \\ 
\end{array}
$
}
\tighten

\begin{document}
\author{B. Dion, L. Marleau and G. Simon}
\address{D\'epartement de Physique, Universit\'e Laval\\
Qu\'ebec, Canada, G1K 7P4}
\title{Leptoquark Pair Production at the Fermilab Tevatron:\\
Signal and Backgrounds}
\date{1996}
\maketitle

\begin{abstract}
We perform a simulation of scalar leptoquark pair production at the Tevatron
($\sqrt{s}=1.8$ TeV and ${\cal L}=100$ pb$^{-1}$) with ISAJET. We also
investigate the dominant sources of Standard Model background: $Z^{*}jj$,
$ZZ $, $WZ$ production and heavy quark $t{\bar t}$. We find that the $Z^{*}jj$
background is dominant. We also evaluate the signal-to-background ratio and
find a discovery reach of 130 GeV (170 GeV) for a branching ratio of $%
BR(LQ\rightarrow eq)=0.5$ ($BR=1$).
\end{abstract}


\section{Introduction}

Symmetry considerations such as family replication, anomaly cancellation and
symmetry between generations are the main motivations behind the prediction
of leptoquarks. They are color-triplet particles which possess baryon number
($B$) and lepton number ($L$). They arise in various extensions of the
Standard Model (SM): GUT's \cite{gut}, Strongly Coupled Standard Model \cite
{af}, composite models \cite{ps}, superstring-inspired models \cite{string}
,where they appear as scalar, vector or even fermionic particles.

Leptoquarks (LQs) could in principle be produced in $e^{+}e^{-}$ \cite{ee}, $%
e\gamma $ \cite{eg}, $ep$ \cite{ep} and hadron colliders \cite{pp,d0,bg,MMS}
. But hadron colliders have two advantages: the production is almost
insensitive to the Yukawa coupling and the available center-of-mass energy ($%
\sqrt{s}$) is much higher. We perform our calculations with the Tevatron in
mind, taking $\sqrt{s}=1.8$ TeV and ${\cal L}=100$ pb$^{-1}$. The current
limits imposed on first generation LQ mass by the D0 collaboration are 133
GeV for $BR(LQ\rightarrow eq)=1$ and 120 GeV for $BR(LQ\rightarrow eq)=0.5$
and $BR(LQ\rightarrow \nu q)=0.5$ \cite{pf}. The object of the present paper
is to evaluate the importance of SM background ($Z^{*}jj$ where the $Z$ is
virtual, $ZZ$, $WZ$ and $t\bar{t}$ events) when producing a pair of scalar
leptoquarks in the context of an $E_6$ model. So far, these major sources of
background have only been identified \cite{bg} but no quantitative analysis
has been made. We implemented leptoquark data into the ISAJET event
generator to achieve this goal.

In this paper, we look at the pair production of first generation $E_6$
scalar leptoquarks with $Q=-1/3$ and evaluate the importance of the SM
sources of background for this signal. In the next section, we first
describe the $E_6$ model. In Section 3, we give the details of our
simulation for the detector and the calorimeter plus the kinematic cuts that
are imposed. The LQ signal and the SM backgrounds ( $Z^{*}jj$, $ZZ$, $WZ$
and heavy quark pair production) are discussed in Section 4. We analyze the
results in Section 5 and conclude in the following section.

\section{Framework}

Up to now, the SM provides a satisfactory description of phenomenology.
However, many problems still remain. Among them, we find gauge hierarchical
and fine tuning problems, no explanation for the existence and the number of
fermion families and too many parameters to be extracted from experiment.

In the context of an $E_6$ model \cite{hr}, the low-energy limit of an $%
E_8\otimes E_8^{^{\prime }}$ heterotic string theory, the gauge hierarchical
and fine tuning problems no longer appear due to the supersymmetric nature
of the theory. In $E_6$ superstring models, each matter supermultiplet lies
in the fundamental ${\bf 27}$ representation which may explain the
replication of fermion families. Another attractive feature of the model
lies in the predictions it makes about low-energy physics. The ${\bf 27}$
representation possess the following $SO(10)$ and $SU(5)$ contents:\newline
\begin{equation}
{\bf 27=[16+10+1]=[10+{\bar{5}}+1]+[5+{\bar{5}}]+[1]}.
\end{equation}
The particles belonging to this representation can be arranged as: 
\begin{equation}
{\bf 27=[(u^c,Q,e^{+}]+(L,d^c)+\nu ^c]+[(D,H)+(D^{*},H^{*})]+N}
\end{equation}
where we have the usual quarks and leptons and their superpartners along
with new particles such as two five-plets $(D,H)$ and $(D^{*},H^{*})$ and an 
$SU(5)$ superfield singlet N. In particular, the superfields $D$ and $D^{*}$
are two $SU(3)$ triplets and $SU(2)$ singlets with electromagnetic charge $%
-1/3$ and $+1/3$ respectively. They possess $B=\pm 1/3$ and $L=\pm 1$.
Contrary to the usual notation, those particles are supersymmetric and we
denote their non-supersymmetric partners as ${\tilde{D}}$ and ${\tilde{D}}
^{*}$. Thus, scalar leptoquarks come out naturally for an $E_6$ model.
Restricting our study to the first generation of fermions, the Yukawa
Lagrangian will take the form:\newline
\begin{equation}
{\cal L}_Y=\lambda _L{\tilde{D}}^{*}\left( e_Lu_L+\nu _Ld_L\right) +\lambda
_R{\tilde{D}}{\bar{e}}_L{\bar{u}}_L+\text{h.c.}  \label{L}
\end{equation}
where $\lambda _L$ and $\lambda _R$ are chosen to be equal to the
electromagnetic charge otherwise there would be important effects in the
process $e^{+}e^{-}\rightarrow q\bar{q}$ \cite{eeqq} arising from the LQ
exchange.

\section{Event simulation}

\subsection{Detector and calorimeter}

We use ISAJET, a Monte-Carlo event simulator, to model the experimental
conditions at the Tevatron. We simulate a toy detector with the following
characteristics for the calorimeter:

\begin{itemize}
\item  cell size: $\bigtriangleup \eta \times \bigtriangleup \phi =0.1\times
0.0875$,

\item  pseudorapidity range: $-4<\eta <4$,

\item  hadronic energy resolution: $70\%/\sqrt{E}$,

\item  electromagnetic energy resolution: $15\%/\sqrt{E}$.
\end{itemize}

\subsection{Kinematic cuts}

In order to qualify as a jet, any hadronic shower must satisfy the following
kinematic cuts:

\begin{itemize}
\item  lie within a cone of radius $R=\sqrt{(\bigtriangleup \eta
)^2+(\bigtriangleup \phi )^2}=0.7$,

\item  have a transverse energy $E_T>25$ GeV,

\item  have a pseudorapidity $|\eta _j|\leq 3.$
\end{itemize}

Similarly, we must impose some cuts on the leptons. More specifically,
electrons are considered isolated if they:

\begin{itemize}
\item  are separated from any jet by $R\geq 0.3$,

\item  have a transverse momentum $p_T>25$ GeV,

\item  have a pseudorapidity $|\eta _l|\leq 2.5.$
\end{itemize}

Our calculations are performed using the PDFLIB distribution functions of
Morfin and Tung (M-T B2) with $\Lambda =191$ GeV \cite{PDFLIB}.

\section{Signal and backgrounds}

\subsection{Leptoquark signal}

We study the pair production of $E_6$ scalar leptoquarks with $Q=-1/3$
decaying into an up quark and an electron or into a down quark and a
electron neutrino with branching ratios of $BR(LQ\rightarrow
eq)=BR(LQ\rightarrow \nu q)=0.5$. For comparison, we also consider the case
of a leptoquark decaying into an up quark and an electron with $%
BR(LQ\rightarrow eq)=1$. In both cases, we assume a Yukawa coupling with $%
k=1 $ in $\alpha _Y=k\alpha _{em}$.

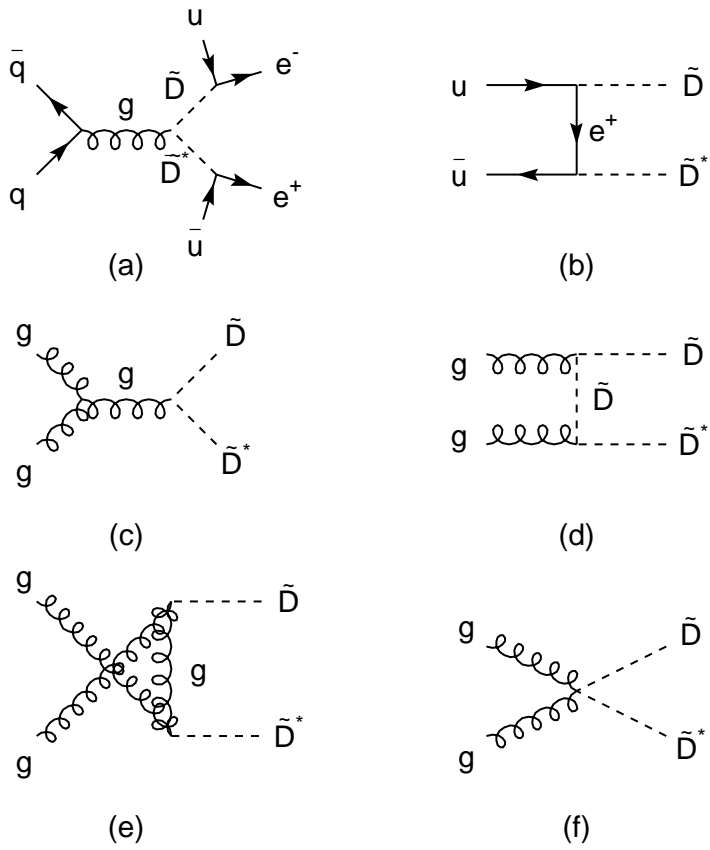
\begin{figure}[tbp]
\caption{Feynman diagrams for leptoquark pair production via ((a), (b)) $q 
\bar{q}$ annihilation and ((c), (d), (e), (f)) gluon fusion.}
\label{feynlq}
\end{figure}

The leptoquark pair production can occur via two different channels, either
via $q\bar{q}$ annihilation $q\bar{q}\rightarrow {\tilde{D}}{\tilde{D}}^{*}$
or gluon fusion $gg\rightarrow {\tilde{D}}{\tilde{D}}^{*}$. The
corresponding Feynman diagrams are shown in Fig. \ref{feynlq}. For the
details of the calculation of the cross section, we refer the reader to Ref. 
\cite{MMS}. We expect the first of these channels to dominate at the
Tevatron.

In the case $BR=0.5$, we obtain three different signals \cite{bg}:

\begin{enumerate}
\begin{enumerate}
\item  2 jets + $e^{+}e^{-}$,

\item  2 jets + ${\not{p}}_T$,

\item  2 jets + $e^{\pm }+{\not{p}}_T$.
\end{enumerate}
\end{enumerate}

The most distinctive of these signals \cite{bg} should be (a) but apart from
that, very little is known about the signal and backgrounds expected from
the SM. The background for signal (b) and (c) should be more important due
to the missing transverse momentum (${\not{p}}_T$). For the purposes of this
paper, we therefore restrict ourselves to 2 jets + $e^{+}e^{-}$. For
simplicity, we impose the same $E_T$ cut values on the jets and the leptons:
25 GeV, 30 GeV, 35 GeV, 40 GeV.

\subsection{SM Backgrounds}

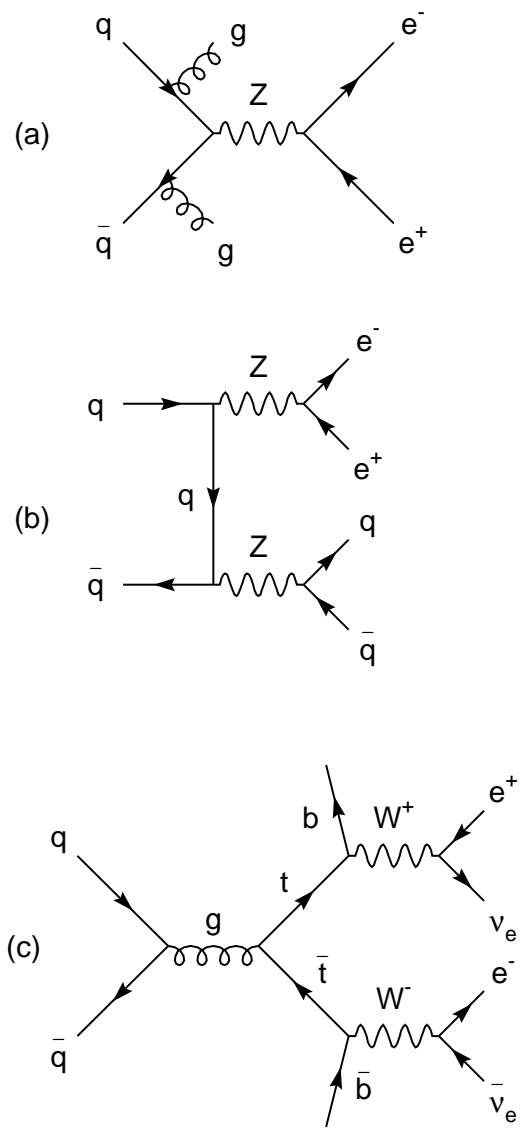
\begin{figure}[tbp]
\caption{Examples of Feynman diagrams for (a) $Z^{*}jj$, (b) $ZZ$ pair and
(c) $t \bar{t}$ production.}
\label{feynbg}
\end{figure}

The most probable sources of background as identified by Refs. \cite{bg} are
(1) $Z^{*}jj$, (2) $ZZ$, $WZ$ and (3) $t\bar{t}$ production (in which the
top is decaying into a $W$ and a $b$ quark). Backgrounds from $b\bar{b}$ and 
$c\bar{ c}$ are ignored. Previous estimates have shown them to be negligible 
\cite{d0}. The Feynman diagrams of the background processes are shown in
Fig. \ref{feynbg}.

\subsubsection{$Z^{*}jj$ background}

The $Z^{*}jj$ background proceed through a Drell-Yan virtual $Z$ (producing
a pair of leptons) along with two jets: $p\bar{p}\rightarrow
Z^{*}jj\rightarrow e^{+}e^{-}jj$ where $j=q,\bar{q},g$. We consider this
background for an invariant mass of the lepton pair ranging from 100 GeV to
260 GeV.

\subsubsection{$ZZ$ and $WZ$ background}

Other possible sources of background come from the production of a pair of $%
Z $ bosons with one of the $Z$ decaying into a lepton pair and the other one
decaying into a pair of quarks and/or the production of a $Z$ and a $W$ with
the $Z$ decaying into a lepton pair and the $W$ decaying into a pair of
quarks. We calculated those backgrounds and found them negligible. The small
magnitude of this background can be explained by the fact that $ZZ$ and $WZ$
productions are suppressed by $\alpha _W^2$ where $\alpha _W$ is the weak
coupling with respect to the LQ process and the $t\bar{t}$ background which
involve only strong interactions and a factor of $\alpha _s^2$ with $\alpha
_s\simeq $ 0.1. Therefore, we will not consider those background processes
any further. In any case, the study of the invariant mass of the pairs could
single them out.

\subsubsection{$t\bar{t}$ background}

The final background process we consider comes from $t\bar{t}$ where the top
is decaying into a bottom, an electron and a neutrino, $\nu _e$. The
presence of neutrinos implies a ${\not{p}}_T$ and a less energetic electron.
However, our analysis has shown that this process can provide some very
energetic jets and thus be an important source of background. We consider $%
M_t$=175 GeV.

\section{Discovering leptoquarks at the Tevatron}

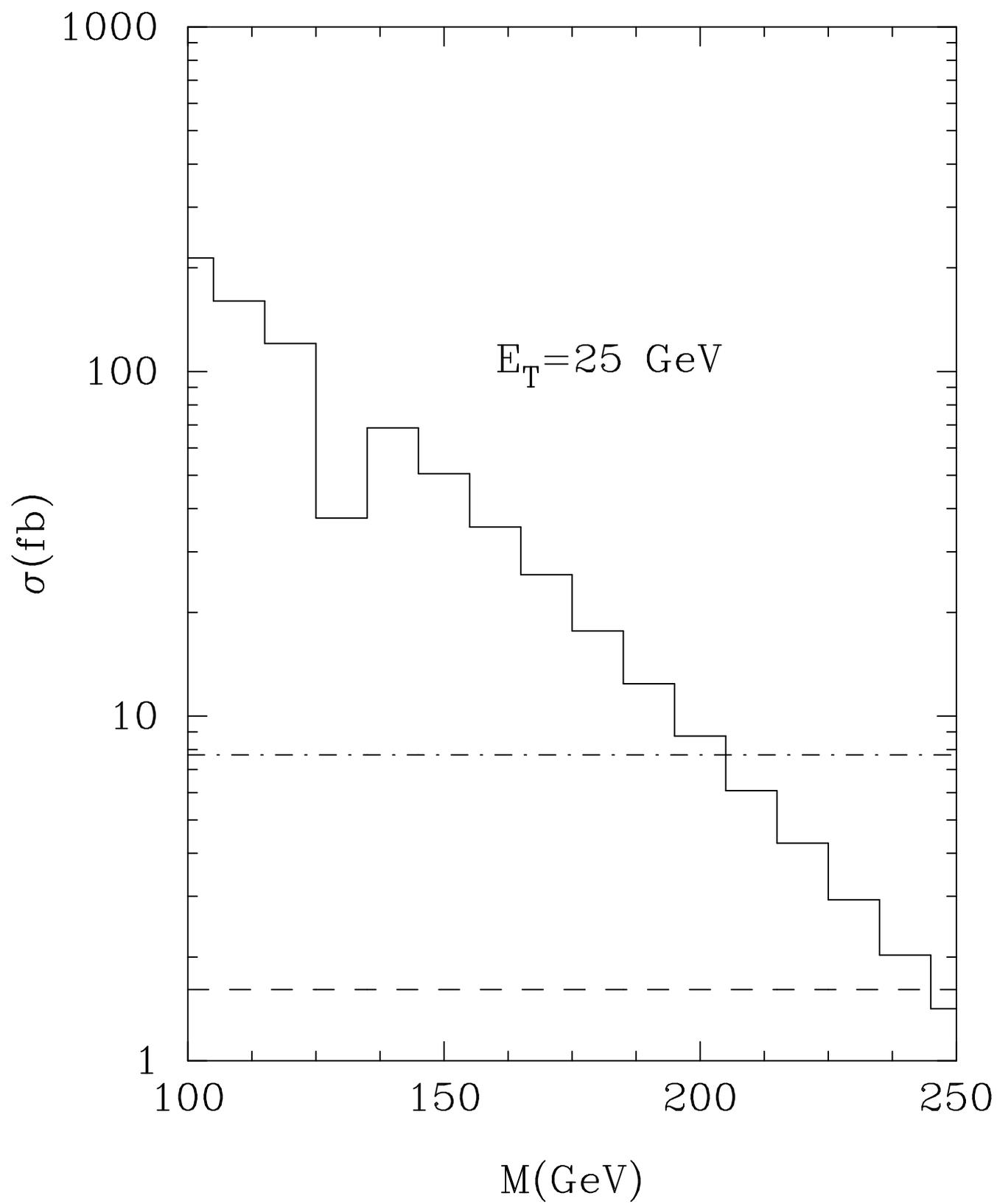
\begin{figure}[tbp]
\caption{Integrated cross section for the production of 2 jets + $e^{+}
e^{-} $ as a funtion of the leptoquark mass for $E_{T}=25$ GeV. The full
line corresponds to the leptoquark signal versus the leptoquark mass ($M_{LQ}
$), the dash-dotted line to the total $Z^{*}jj$ background and the dashed
line to $t{\bar t}$ background for $M_{t}$=175 GeV.}
\label{ET25}
\end{figure}

In Fig. \ref{ET25}, we first show the results for the total cross section
for the leptoquark signal (2 jets + $e^{+}e^{-}$) as a function of the
leptoquark mass, $M_{LQ}$ (solid lines) . For comparison, we also show the $t%
\bar{t}$ background (dashed lines) evaluated at $M_t$=175 GeV as well as the 
$Z^{*}jj$ background (dot-dashed lines) integrated over an invariant mass of
the lepton pair ranging from 100 GeV to 260 GeV. The $Z^{*}jj$ signal
dominates the background while the $t\bar{t}$ signal is relatively small.
Requiring a statistical significance $\geq 5\sigma $, where $\sigma
=N_{signal}/\sqrt{N_{bckg}}$, we find that the leptoquark signal dominates
the background for leptoquark masses up to 150 GeV (200 GeV) for $BR=0.5$ ($%
BR=1$). These ranges are found to be insensitive to the value of $E_T$ as
long as it remains in the region 25 GeV $\leq E_T\leq $ 40 GeV.

The $t\bar{t}$ background is higher than the heavy quark background
predicted by D0 \cite{d0} coming from $c\bar{c}$ and $b\bar{b}$ due to the
masses of the particles involved. The relevance of this background can be
associated to the available transverse energy which is comparable to that in
the leptoquark process analyzed in this work. The absence of missing ${\not%
{p}}_T$ is however characteristic in the LQ case. In fact, such a cut
increases the signal-to-background ratio even more but reduces the signal by
about 10\%. However, since the $t\bar{t}$ background is not significant
here, we chose not to include this cut in the rest of the analysis.

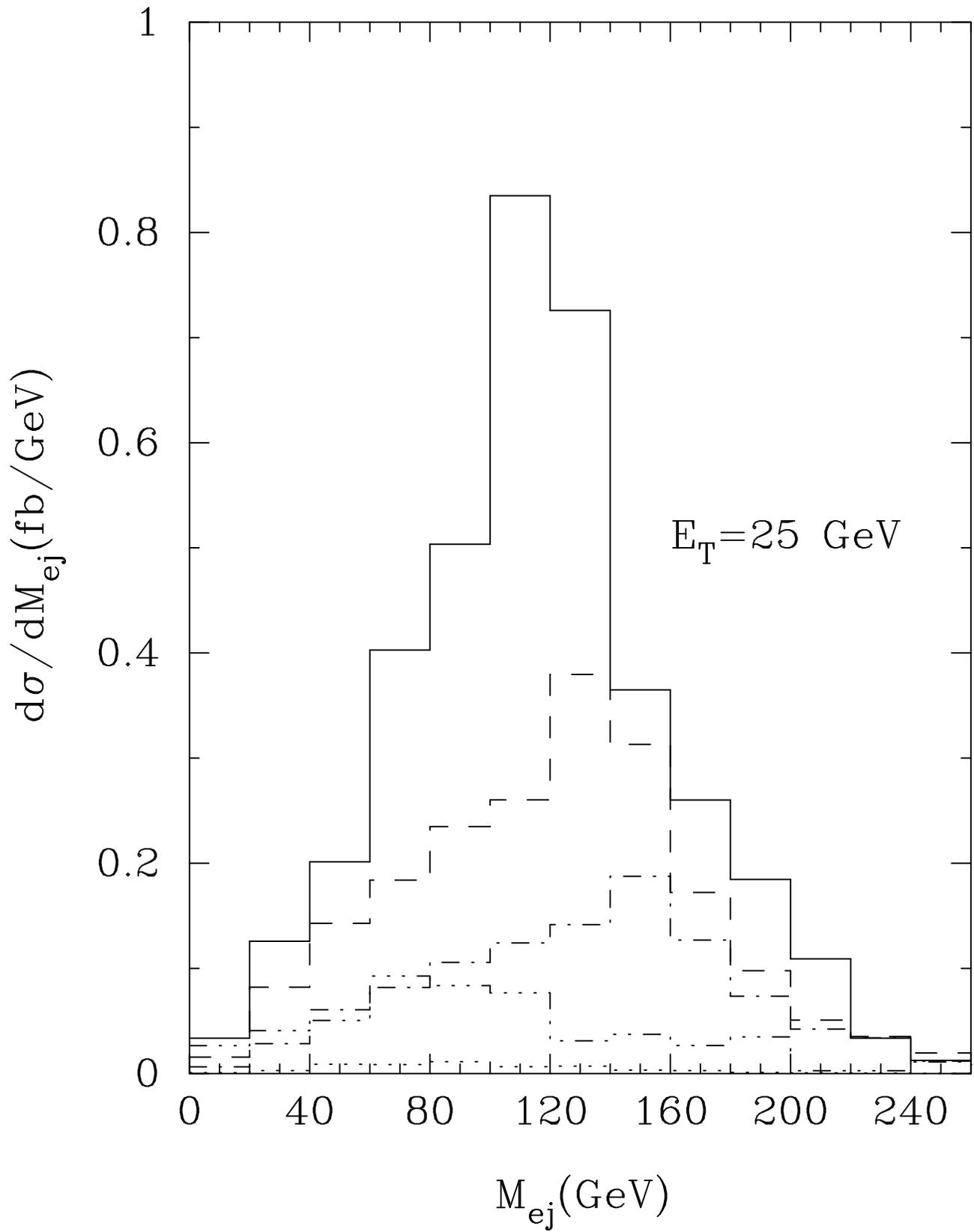
\begin{figure}[tbp]
\caption{Distribution of the invariant mass of the lepton-jet pair for the
production of 2 jets + $e^{+} e^{-}$ for $E_{T}=25$ GeV. The solid, dashed
and dashed-dotted lines correspond to the leptoquark signal with $M_{LQ}$%
=130, 150 and 170 GeV respectively. The dash-dot-dot-dotted lines correspond
to the $Z^{*}jj$ background and the dotted lines to $t{\bar t}$ background
(at the bottom of the plot).}
\label{minvet25}
\end{figure}

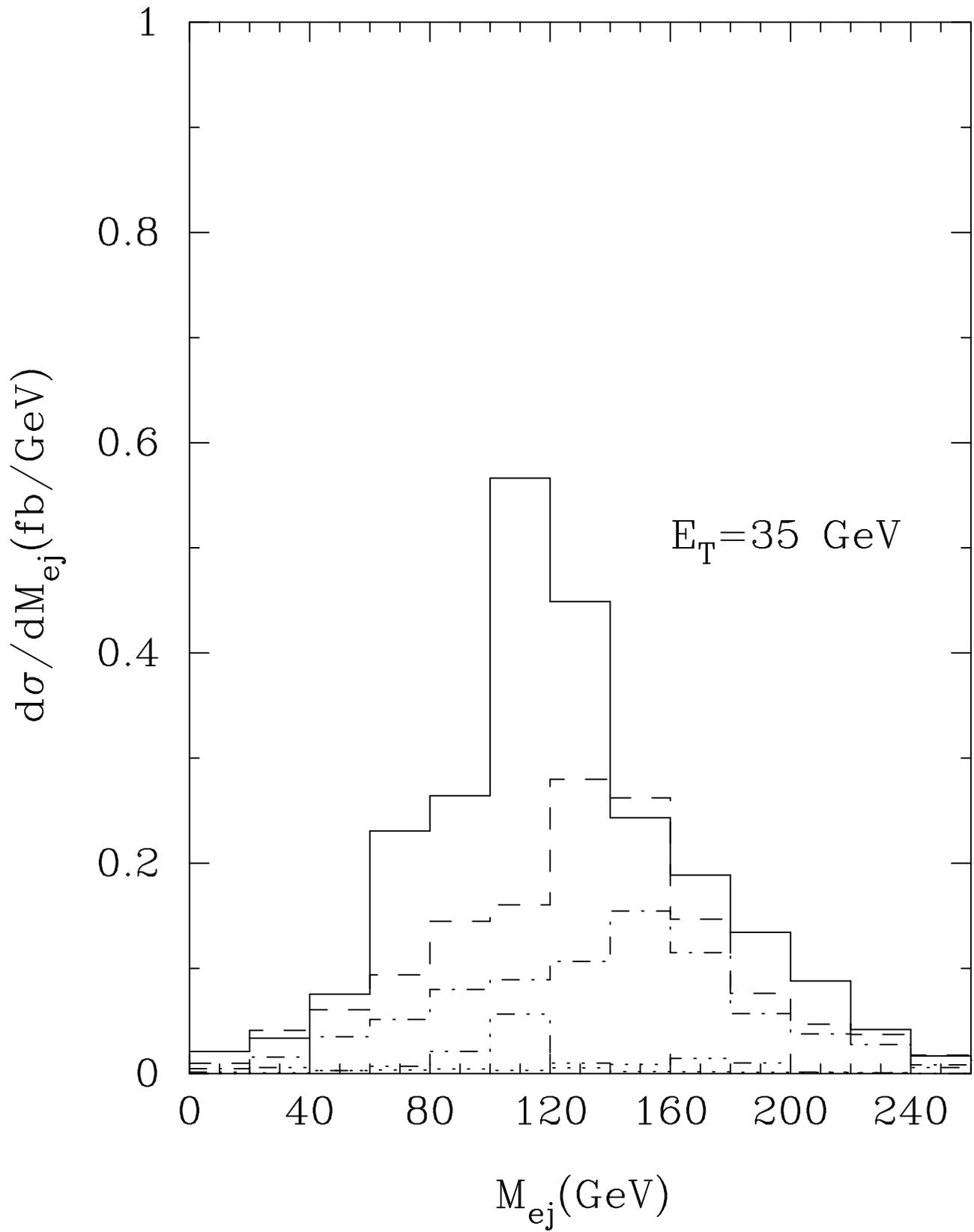
\begin{figure}[tbp]
\caption{Same as Fig. 3 but for $E_{T}=35$ GeV.}
\label{minvet35}
\end{figure}

In general, one expects that in leptoquark production there will be a strong
lepton-jet correlation due to the leptoquark decay while such correlation
should be absent in $t\bar{t}$ and $Z^{*}jj$ backgrounds. Indeed, even if
the signal and background total cross-section were comparable in magnitude,
there can still be a detectable signal in the form of a peak in the
invariant mass distribution of the lepton-jet pairs. The distribution in the
invariant mass of the lepton-jet pairs are presented in Figs. \ref{minvet25}-%
\ref{minvet35} for the various processes for $E_T=25$ GeV and $35$ GeV
respectively. The invariant mass $M_{ej}$ was calculated by pairing the most
energetic electron with the least energetic jet. This particular choice is
based on the reasonable assumption that the two leptoquarks will emerge with
approximately the same energy and that this lepton-jet pair will correspond
to the decay products of the same leptoquark. Each figure displays the
leptoquark signal for leptoquark mass inputs of $M_{LQ}=130$ GeV (solid
lines), $M_{LQ}=150$ GeV (dashed lines) and $M_{LQ}=170$ GeV (dash-dotted
lines) as well as the background due to the $Z^{*}jj$ (dash-dot-dot-dotted
lines) and the $t\bar{t}$ (dotted lines) processes. Here, we have applied an
invariant mass cut on the lepton pairs with $81$ GeV $\leq
M_{e^{+}e^{-}}\leq 101$ GeV in order to eliminate the events near the $Z^0$
peak and minimize the $Z^{*}jj$ background. We see from Figs. \ref{minvet25}-%
\ref{minvet35} that the optimal $E_T$ cut is $25$ GeV. An $E_T$ cut of $35$
GeV reduces the background significantly. Our results show the expected
strong lepton-jet correlation in the leptoquark signal. The $Z^{*}jj$
background is mostly concentrated between $50\leq M_{ej}\leq 100$ GeV but
remains the dominant background for all values of the invariant mass. The $%
5\sigma $ statistical significance is satisfied for $M_{LQ}=130$ and $150$
GeV with an $E_T$ cut of $25$ GeV.

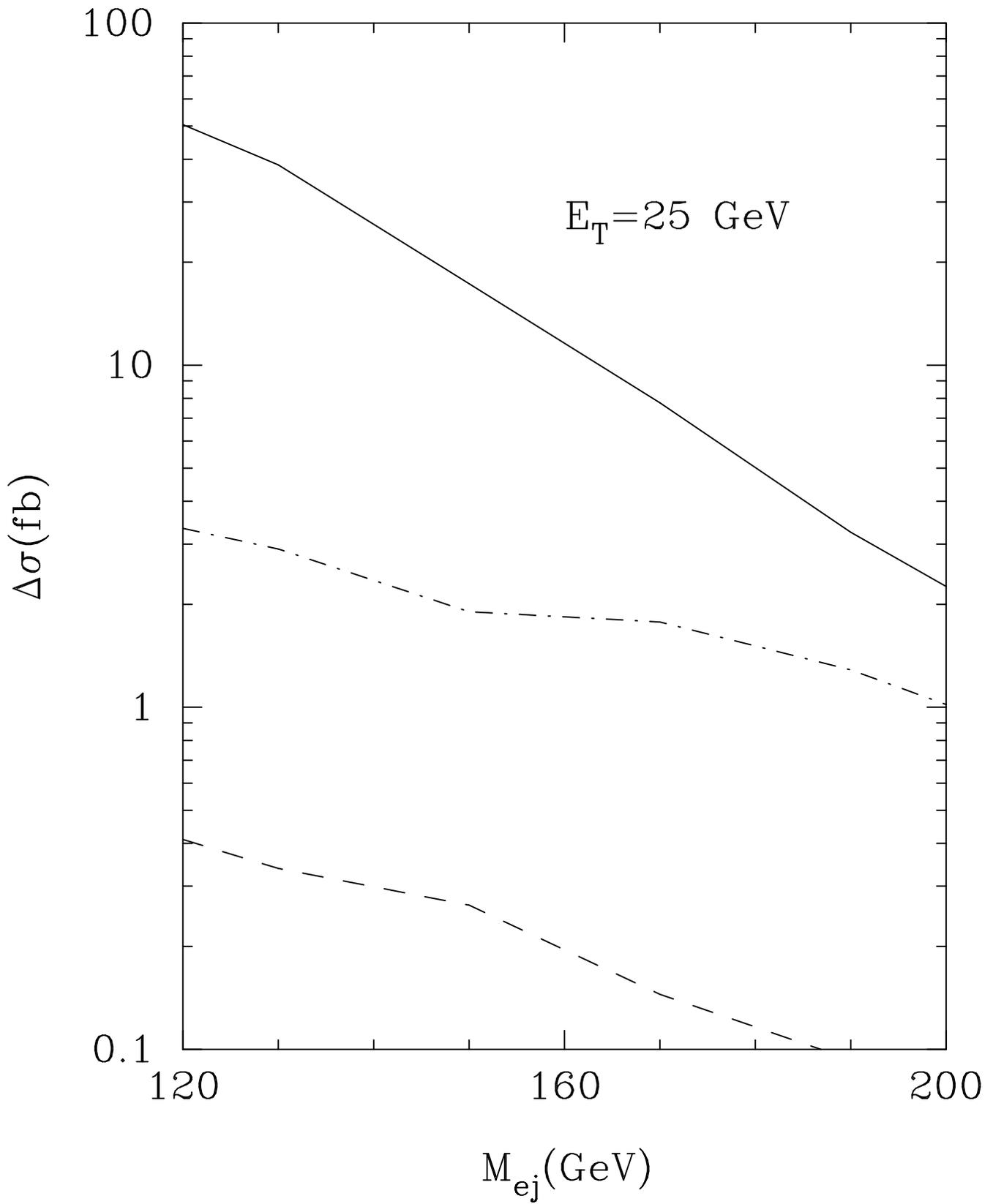
\begin{figure}[tbp]
\caption{Partial cross section within a bin of width $\Delta M_{ej}$=60 GeV
around $M_{ej}=M_{LQ}$ as a function of the invariant mass of the
electron-jet pair for $E_{T}=25$ GeV.}
\label{deltasigma}
\end{figure}

In order to estimate the relative importance of the signal to the background
near the peak in the $M_{ej}\,$distribution, we calculate the partial cross
section $\Delta \sigma $ within a bin of width $\Delta M_{ej}=60$ GeV around 
$M_{ej}=M_{LQ}$ as a function of the invariant mass of the electron-jet pair
for $E_T=25$ GeV. The calculations repeated for several intermediate values
of $M_{LQ}$ and are shown in Fig. \ref{deltasigma}. We can then see more
clearly how that the signal-to-background ratio is affected by the
leptoquark mass. Also, we find that the $5\sigma $ statistical significance
condition is satisfied for $M_{LQ}\leq 150$ GeV.

Coming back to figure \ref{ET25}, we can estimate the discovery reach of
scalar leptoquarks at the Tevatron. To that effect, we require a minimum of
10 events in addition to the $5\sigma $ statistical significance condition
imposed on the background. The integrated luminosity for the Tevatron
current run is 100 pb$^{-1}$ which leads to a discovery mass limit of $130$
GeV for $E_T=25$ GeV (Fig. \ref{ET25}). A similar estimate can be easily
carried for leptoquarks that decay with $BR(LQ\rightarrow eq)=1$ by arguing
that the cross section for the production of 2 jets + $e^{+}e^{-}$ is four
times larger in this case. Accordingly, the discovery reach is increased up
to $170$ GeV. On the other hand, the proposed luminosity upgrades of the
Tevatron should enhance significantly the signal and increase the discovery
reach to approximately $200$ GeV for $BR=0.5$ (with a luminosity 2 fb$^{-1}$%
).

Summarizing, we carried out a simulation for first-generation scalar
leptoquark pair production within the context of an $E_6$ model taking into
account the SM backgrounds for the most promising signal, i.e. 2 jets + $%
e^{+}e^{-}$. The results are found to be sensitive on the choice of $E_T$
cut for the jets and while background is dominated by the $Z^{*}jj$ process,
it does not overwhelm the signal for leptoquarks at the Tevatron.  We find a
discovery mass limit of 130 GeV ($BR=0.5$) and 170 GeV ($BR=1$) for an
optimum $E_T$ cut of 25 GeV.

\acknowledgements
G. Simon would like to thank F. Paige for his help during the installation
of ISAJET. This research was supported by the Natural Sciences and
Engineering Research Council of Canada and by the Fonds pour la Formation de
Chercheurs et l'Aide \`{a} la Recherche du Qu\'{e}bec.

\end{document}